\documentclass[pra,superscriptaddress,showpacs,twocolumn,10pt]{revtex4}
\usepackage{amssymb}
\usepackage{graphicx}

\def\>{\rangle}

\begin{document}
\title{When catalysis is useful for probabilistic entanglement transformation}
\author{Yuan Feng}
\email{fengy99g@mails.tsinghua.edu.cn}
\author{Runyao Duan}
\email{dry02@mails.tsinghua.edu.cn}
\author{Mingsheng Ying}
\email{yingmsh@tsinghua.edu.cn} \affiliation{State Key Laboratory
of Intelligent Technology and Systems, Department of Computer
Science and Technology Tsinghua University, Beijing, China,
100084}

\begin{abstract}
We determine all $2\times 2$ quantum states that can serve as
useful catalysts for a given probabilistic entanglement
transformation, in the sense that they can increase the maximal
transformation probability. When higher-dimensional catalysts are
considered, a sufficient and necessary condition is derived under
which a certain probabilistic transformation has useful catalysts.
\end{abstract}
 \maketitle

In the field of quantum information theory, entanglement plays an
essential role in quantum information processing such as quantum
cryptography \cite{BB84}, quantum superdense coding \cite{BS92}
and quantum teleportation \cite{BBC+93}. When entanglement is
treated as a type of resource, the study of transformations
between different forms of entanglement becomes very crucial. It
is well known that the entanglement quantity shared among separate
parties cannot be increased only using local operations on the
separate subsystems and classical communication between them (or
LOCC for short). The restriction on the possibility of
entanglement transformations that can be realized by LOCC is,
however, beyond this. Nielsen proved in his brilliant work
\cite{NI99} that a pure bipartite entangled quantum state
$|\psi_1\rangle$ can be transformed into another pure bipartite
entangled state $|\psi_2\rangle$ by LOCC if and only if
$\lambda_{\psi_1}\prec\lambda_{\psi_2}$, where the probability
vectors $\lambda_{\psi_1}$ and $\lambda_{\psi_2}$ denote the
Schmidt coefficient vectors of $|\psi_1\rangle$ and
$|\psi_2\rangle$, respectively. Here the symbol $\prec$ stands for
the ``majorization relation". An $n$-dimensional probability
vector $x$ is said to be majorized by another $n$-dimensional
probability vector $y$, denoted by $x\prec y$, if the following
relations hold
$$\sum_{i=1}^l x^\downarrow_i \leq \sum_{i=1}^l y^\downarrow_i {\rm \ \ \ for\ any\ \ \ }1\leq l< n,$$
where $x^\downarrow$ denotes the vector obtained by rearranging
the components of $x$ in nonincreasing order.

What Nielsen has done indeed gives a connection between the theory
of majorization in linear algebra \cite{MO79} and the entanglement
transformation. Furthermore, since the sufficient and necessary
condition is very easy to check, it is extremely useful to decide
whether one pure bipartite entangled state can be transformed into
another pure bipartite state by LOCC. There exist, however,
incomparable states in the sense that any one cannot be
transformed into another only using LOCC. To cope with the
transformation between incomparable states, Vidal \cite{Vidal99}
generalized Nielsen's work with a probabilistic manner. He found
that although a deterministic transformation cannot be realized
between incomparable states, a probabilistic one is always
possible. Furthermore, he gave an explicit expression of the
maximal probability of transforming one state to another. To be
more specific, let $P(|\psi\rangle \rightarrow |\phi\rangle)$
denote the maximal transformation probability of transforming
$|\psi\rangle$ into $|\phi\rangle$ by LOCC; then
$$P(|\psi\rangle \rightarrow |\phi\rangle)=\min_{1\leq
l\leq n} \frac{E_l(\lambda_\psi)}{E_l(\lambda_\phi)},$$ where $n$
is the maximum of the Schmidt numbers of $|\psi\rangle$ and
$|\phi\rangle$, and $E_l(x)$ denotes the abbreviation of
$\sum_{i=l}^n x^\downarrow_i$ for probability vector $x$.

Another interesting phenomenon was discovered by Jonathan and
Plenio \cite{JP99} that sometimes an entangled state can help in
making impossible entanglement transformations into possible
without being consumed at all. That is, there exist quantum states
$|\psi_1\rangle$, $|\psi_2\rangle$, and $|\phi\rangle$ such that
$|\psi_1\rangle\nrightarrow |\psi_2\rangle$ but
$|\psi_1\rangle\otimes|\phi\rangle \rightarrow
|\psi_2\rangle\otimes|\phi\rangle$. In this transformation, the
role of the state $|\phi\rangle$ is just like a catalyst in a
chemical process. They found by examining an example that in some
cases an appropriately chosen catalyst can increase the maximal
transformation probability of incomparable states. It was also
shown that enhancement of the maximal transformation probability
is not always possible. However, there were no further results
about such an interesting field in their paper.

In this paper, we examine the ability of catalysts in a
probabilistic entanglement transformation. We first consider the
simple case of when a given probabilistic transformation has
useful $2\times 2$ catalysts and determine all of them. Then a
sufficient and necessary condition is derive which can decide
whether or not a certain transformation has (not necessarily
$2\times 2$) useful catalysts.

For simplicity, in what follows we denote a quantum state by the
probability vector of its Schmidt coefficients. This will not
cause any confusion because it is well known that the fundamental
properties of a bipartite quantum state under LOCC are completely
determined by its Schmidt coefficients. Therefore, from now on, we
consider only probability vectors instead of quantum states and
always identify a probability vector with the quantum state
represented by it.

Suppose $x,y$ are two $n$-dimensional probability vectors and the
components are arranged nonincreasingly. It is well known that if
$P(x\rightarrow y)=E_n(x)/E_n(y)$  or $P(x\rightarrow
y)=E_1(x)/E_1(y)(=1)$, then the maximal probability of
transforming $x$ into $y$ cannot be increased by any catalyst.
That is, for any probability vector $c$, we have $P(x\otimes
c\rightarrow y \otimes c)=P(x\rightarrow y)$. Thus, in what
follows, we assume
\begin{equation}\label{assum} P(x\rightarrow y)
<\frac{E_n(x)}{E_n(y)}\ \ {,\ \ }P(x\rightarrow y)
<\frac{E_1(x)}{E_1(y)}.\end{equation} Without loss of generality,
we concentrate on catalysts with nonzero components, since $c$ and
$c\oplus 0$ have the same catalysis ability for any probability
vector $c$ in the sense that in any situation, if one serves as a
partial catalyst for some transformation, so does the other for
the same transformation. Let
$$L=\{l:1< l< n {\ \ \rm and\ \ \ }
P(x\rightarrow y)=\frac{E_l(x)}{E_l(y)}\}.$$ In what follows, we
derive a sufficient and necessary condition when a two-dimensional
catalyst can increase the maximal transformation probability from
$x$ to $y$. In order to state the theorem compactly, we first
denote
\begin{equation} \label{equ:m}
m_{r_1}^{r_2}=
\min\{\frac{x_{r_1-1}}{x_{r_2}},\frac{y_{r_1-1}}{y_{r_2}}\}
\end{equation}
and
\begin{equation}\label{equ:M}
M_{r_1}^{r_2}=
\max\{\frac{x_{r_1}}{x_{r_2-1}},\frac{y_{r_1}}{y_{r_2-1}}\}
\end{equation}
for any $r_1, r_2\in L$. Furthermore, we let $M_{n+1}^{r_2}=0$.

{\bf Theorem 1.} The maximal probability of transforming $x$ into
$y$ can be increased by a 2-dimensional catalyst if and only if
the set
\begin{equation}\label{equ:S}
S= \bigcap\{(0,M_{r_1}^{r_2})\cup(m_{r_1}^{r_2},1)\}
\end{equation}
is not empty, where the intersection is taken over all pairs of
$r_1,r_2$ such that $r_1,r_2\in L\cup\{n+1\}$, $r_1\geq r_2$, and
$r_2<n+1$. In fact, any two-dimensional probability vector
$(c_1,c_2)$ with $c_1\geq c_2$ can serve as a useful catalyst for
transforming $x$ into $y$ if and only if $c_2/c_1\in S$.

{\bf Proof.} Suppose $c=(c_1,c_2)$ is a $2$-dimensional
probability vector and $c_1\geq c_2$. We need only show that $c$
cannot serve as a useful catalyst for transforming $x$ into $y$,
that is,
$$P(x\otimes c\rightarrow y\otimes c)=P(x\rightarrow y)$$ if and
only if there exist $r_1,r_2\in L\cup \{n+1\}$, $r_1\geq r_2$, and
$r_2<n+1$, such that
\begin{equation}\label{conds}
M_{r_1}^{r_2}\leq \frac{c_2}{c_1}\leq m_{r_1}^{r_2},
\end{equation} where any constraint having meaningless terms is satisfied automatically.

For an arbitrarily fixed integer $l$ satisfying $1<l\leq 2n$, we
can arrange the summands in $E_l(x\otimes c)$ as
\begin{equation}\label{equ:el}
E_l(x\otimes c)=c_1\sum_{i=r_1}^n x_i + c_2\sum_{i=r_2}^n x_i.
\end{equation}
Here $r_i$, $1\leq r_i\leq n+1$, denotes the smallest index of the
components of $x$ in the summands of $E_l(x\otimes c)$ that have
the form $c_ix_j$, where $1\leq j\leq n$. The case $r_i=n+1$
denotes that any term that has the form $c_ix_j$ does not occur.
In the case of repeated values of components of $x\otimes c$, we
regard terms with larger $i$, and larger $j$ if they have the same
$i$, to be included in the sum first.

From these assumptions, we can show that $r_1\geq r_2$. Otherwise,
$r_1\leq r_2-1$, and from the fact $c_1 x_{r_1}$ is in the
summands of $E_l(x\otimes c)$ while $c_2 x_{r_2-1}$ is not, we can
deduce that $c_1 x_{r_1}\leq c_2 x_{r_2-1}$. But on the other
hand, we have $c_1 \geq c_2 $ and $x_{r_1}\geq x_{r_2-1}$. So it
follows that $c_1 = c_2 $ and $x_{r_1}= x_{r_2-1}$. Especially,
$c_1 x_{r_2-1}=c_2x_{r_2-1}$ which contradicts our assumption that
the term with larger $i$ is included in $E_l(x\otimes c)$ first,
since the former is while the latter is not included in $s_x$.
Furthermore, from Eq. (\ref{equ:el}) we have $r_1+r_2=l+1$; then,
$r_2<n+1$ since $1< l\leq 2n$.

Now, by the definition of $E_l(y\otimes c)$ and $P(x\rightarrow
y)$, the following inequality is easy to check:
\begin{equation}
\begin{array}{rl}
\frac{\displaystyle E_l(x\otimes c)}{\displaystyle E_l(y\otimes
c)}&\geq \frac{\displaystyle c_1\sum_{i=r_1}^n x_i +
c_2\sum_{i=r_2}^n
x_i}{\displaystyle c_1\sum_{i=r_1}^n y_i + c_2\sum_{i=r_2}^n y_i}\\
\\
& \geq \frac{\displaystyle P(x\rightarrow y)(c_1\sum_{i=r_1}^n y_i
+ c_2\sum_{i=r_2}^n y_i)}{\displaystyle c_1\sum_{i=r_1}^n y_i +
c_2\sum_{i=r_2}^n
y_i}\\
\\
&=P(x\rightarrow y).
\end{array}
\end{equation}
The first equality holds if and only if $E_l(y\otimes
c)=c_1\sum_{i=r_1}^n y_i + c_2\sum_{i=r_2}^n y_i$ while the second
equality holds if and only if $r_1$ and $r_2$ are both included in
$L$ or, $r_2\in L$ and $r_1=n+1$. Notice that $l$ and $r_i$ can be
uniquely determined by each other from Eq. (\ref{equ:el}); it
follows that the sufficient and necessary condition of when
$P(x\otimes c\rightarrow y\otimes c)=P(x\rightarrow y)$ is there
exist $r_1,r_2\in L\cup\{n+1\}$ satisfying $r_1\geq r_2$ and
$r_2<n+1$, such that
\begin{equation}\label{condx}
E_l(x\otimes c)= c_1\sum_{i=r_1}^n x_i + c_2\sum_{i=r_2}^n x_i
\end{equation}
and
\begin{equation}\label{condy}
E_l(y\otimes c)= c_1\sum_{i=r_1}^n y_i + c_2\sum_{i=r_2}^n y_i.
\end{equation}

In what follows, we derive the conditions presented in Eq.
(\ref{conds}) from Eqs.(\ref{condx}) and (\ref{condy}). In fact,
what Eq. (\ref{condx}) says is simply that $c_1x_{r_1-1}\geq
c_2x_{r_2}$ and $c_2x_{r_2-1}\geq c_1x_{r_1}$ or, equivalently,
\begin{equation}\label{condx1}
\frac{x_{r_1}}{x_{r_2-1}}\leq \frac{c_2}{c_1}\leq
\frac{x_{r_1-1}}{x_{r_2}}.
\end{equation}
The special case when $r_1$ takes value $n+1$ can be included in
Eq. (\ref{condx1}) simply by assuming that the constraints in Eq.
(\ref{condx1}) containing meaningless terms are automatically
satisfied. Analogously, we can show that Eq. (\ref{condy}) is
equivalent to
\begin{equation}\label{condy1}
\frac{y_{r_1}}{y_{r_2-1}}\leq \frac{c_2}{c_1}\leq
\frac{y_{r_1-1}}{y_{r_2}}.
\end{equation}
Combining Eqs.(\ref{condx1}) and (\ref{condy1}) together and
noticing the denotations in Eqs.(\ref{equ:m}) and (\ref{equ:M}),
we derive the sufficient and necessary condition for
two-dimensional probability vector $c$ such that $P(x\otimes
c\rightarrow y\otimes c)=P(x\rightarrow y)$ is just what Eq.
(\ref{conds}) presents. That completes our proof. \hfill $\square$

A special and perhaps more interesting case of the above theorem
is when the number of elements in $L$ is 1, that is $L=\{l\}$ for
some $1<l< n$. In this case, the possible values of the pair
$(r_1, r_2)$ are just $(l,l)$ and $(n+1,l)$. So the set $S$ in Eq.
(\ref{equ:S}) is simply $(0,M_l^l)\cap (m_{n+1}^l,1)$ and the
sufficient and necessary condition of when two-dimensional
catalysts exist which can increase the maximal probability of
transforming $x$ into $y$ is $m_{n+1}^l < M_l^l$, that is,
\begin{equation}\label{existconds}
\min\{\frac{x_{n}}{x_{l}},\frac{y_{n}}{y_l}\} <
\max\{\frac{x_l}{x_{l-1}},\frac{y_l}{y_{l-1}}\}.
\end{equation}

Furthermore, any two-dimensional probability vector $c$, $c_1\geq
c_2$, which satisfies
\begin{equation}\label{2conds}
\min\{\frac{x_{n}}{x_{l}},\frac{y_{n}}{y_l}\} < \frac{c_2}{c_1}<
\max\{\frac{x_l}{x_{l-1}},\frac{y_l}{y_{l-1}}\},
\end{equation}
can be a useful catalyst for this transformation. In the simplest
case of $n=3$ (notice that when $n=2$, any entanglement
transformation has no catalyst), the set $L$ must be $\{2\}$ and
furthermore, from the assumption Eq. (\ref{assum}) we have
$x_3/x_2>y_3/y_2$ and $x_2/x_1<y_2/y_1$. It follows that when $x$
and $y$ are both three-dimensional, the sufficient and necessary
condition for them to have a useful two-dimensional catalyst is
\begin{equation}
\frac{y_{3}}{y_2} < \frac{y_2}{y_{1}}
\end{equation}
and any $c=(c_1,c_2)$ with $c_1\geq c_2$ and
\begin{equation}\label{existcond3}
\frac{y_{3}}{y_2} < \frac{c_2}{c_1}< \frac{y_2}{y_{1}}
\end{equation} can
increase the maximal transformation probability. Note that these
conditions are all irrelevant to $x$.

To illustrate the utility of the above theorem, let us give some
simple examples.

\vspace{1em}

{\bf Example 1.} This example is given by Jonathan and Plenio in
\cite{JP99}. Let $x=(0.6,0.2,0.2)$ and $y=(0.5,0.4,0.1)$, we have
$y_{3}/y_2=0.25$ and $y_2/y_{1}=0.8.$ So from Eq.
(\ref{existcond3}), any state $c=(c_1,c_2)$, $c_1\geq c_2$, can
serve as a useful catalyst for transforming $x$ into $y$, provided
that $0.25<c_2/c_1<0.8$ or, equivalently, $5/9<c_1<4/5$.
Especially, when choosing $c_1=0.65$, we get $c=(0.65,0.35)$,
which is the one given in \cite{JP99}.

Suppose $x$ is just as above while $y=(0.5,0.3,0.2)$; then,
$y_{3}/y_2=2/3$ and $y_2/y_{1}=0.6.$ Since $0.6<2/3$, we deduce
that any two-dimensional state cannot serve as a useful catalyst
for the probabilistic transformation from $x$ to $y$ in the sense
that it cannot increase the maximal transformation probability.

\vspace{1em}

{\bf Example 2.} This well-known example is exactly the original
one that Jonathan and Plenio used to demonstrate entanglement
catalysis \cite{JP99}. Let $x=(0.4, 0.4, 0.1, 0.1)$ and $y=(0.5,
0.25, 0.25, 0)$; then $L=\{3\}$, and from (\ref{existconds}) we
have
$$\min\{\frac{x_{4}}{x_{3}},\frac{y_{4}}{y_3}\}=\min\{1,0\}=0$$
and
$$\max\{\frac{x_3}{x_{2}},\frac{y_3}{y_{2}}\}=\max\{0.25,1\}=1.$$
It follows from Eq. (\ref{2conds}) that any state $c=(c_1,c_2)$,
$0<c_2/c_1<1$, can serve as a useful catalyst. That is, any
two-dimensional nonpure and nonuniform state can increase the
maximal transformation probability from $x$ to $y$.

\vspace{1em}

We have examined when there exists a two-dimensional catalyst
which is useful for probabilistic transformation. In what follows,
we consider the case of higher-dimensional catalysts and derive a
sufficient and necessary condition for a certain probabilistic
transformation to have a useful (not necessarily two-dimensional)
catalyst. More important, the proof process indeed constructs an
appropriate catalyst explicitly. Some techniques in the proof are
from Lemma 4 in \cite{DK01}.

\vspace{1em}

{\bf Theorem 2.} Suppose $x$ and $y$ are two $n$-dimensional
probability vectors with the components ordered nonincreasingly.
Then there exists a probability vector $c$ such that $P(x\otimes
c\rightarrow y\otimes c)>P(x\rightarrow y)$ if and only if
$$P(x\rightarrow y)<\min\{x_n/y_n,1\}.$$

{\bf Proof.} The ``only if" part is easy and we omit the details
here. The proof of ``if" is as follows.

We denote $P(x\rightarrow y)$ as $P$ for simplicity in this proof.
Let $h$, $1\leq h<n$, be the smallest index of the components such
that $x_h/y_h\not =P$ and $\alpha$ be a positive real number such
that $Py_n/x_n<\alpha<1$. Furthermore, if $P>x_h/y_h$, then assume
$\alpha >x_h/(Py_h)$; otherwise, assume $\alpha >Py_h/x_h$. Let
$k$ be a positive integer such that $x_n>x_h\alpha^{k-1}$  and
$$c=(1,\alpha,\alpha^2,\dots,\alpha^{k-1}).$$
We omit the normalization of $c$ here. In what follows, we show
that for any $1<l<nk$, $E_l(x\otimes c)>PE_l(y\otimes c)$; then,
the catalyst we constructed above indeed increases the maximal
transformation probability.

Fix $l$ as an arbitrary integer that satisfies $1< l< nk$ and
denote $s_x=E_l(x\otimes c)$. It is obvious that we can arrange
the summands in $s_x$ as
$$s_x=\sum_{i=1}^n\sum_{j=r_i}^{k-1} x_i\alpha^j.$$
Here $r_i$, $0\leq r_i\leq k$, denotes the minimal power of
$\alpha$ of the terms in the summands of $s_x$ that have the form
$x_i\alpha^j$, where $1\leq i\leq n$ and $0\leq j\leq k-1$. The
case $r_i=k$ denotes that any term that has the form $x_i\alpha^j$
does not occur. Again, we regard terms with larger $i$ to be
included in the sum first in the case of repeated values of
components of $x\otimes c$. Consider the sum
$$s_y=P\sum_{i=1}^n\sum_{j=r_i}^{k-1} y_i\alpha^j.$$
It is obvious that $s_y\geq PE_l(y\otimes c)$ by definition. On
the other hand, we can rearrange the summands of $s_x$ and $s_y$
,respectively, as
$$s_x=\sum_{j=1}^{k-1}\alpha^j\sum_{i=t_j}^n x_i {\rm\ \ and
\ \ }s_y=\sum_{j=1}^{k-1}\alpha^jP\sum_{i=t_j}^n y_i,$$ where
$1\leq t_j\leq n+1$. Since
$$\sum_{i=t_j}^n x_i=E_{t_j}(x)\geq PE_{t_j}(y)=P\sum_{i=t_j}^n
y_i$$ by the definition of $P$, we have $s_x\geq s_y$. Now, if
$s_x>s_y$, then $s_x>PE_l(y\otimes c)$. In the case of $s_x=s_y$,
let $m_y$ denote the minimum of the components of $Py\otimes c$
not included in $s_y$ and $M_y$ denote the maximum included in
$s_y$. If we can prove $m_y<M_y$, then by swapping $m_y$ and $M_y$
(that is, including $m_y$ to and excluding $M_y$ from $s_y$), we
can show $s_x>PE_l(y\otimes c)$ again and that will complete the
proof of this theorem.

In what follows, we prove that in the assumption of $s_x=s_y$,
$m_y<M_y$ holds. Suppose on the contrary $m_y\geq M_y$; then,
$s_y=PE_l(y\otimes c)$. It is then not difficult to show $m_x\geq
m_y$ and $M_x\leq M_y$, where $m_x$ and $M_x$ are defined
analogously to $m_y$ and $M_y$, since $m_x<m_y$ leads to
$$E_{l-1}(x\otimes c)=s_x+m_x<s_y+m_y=PE_{l-1}(y\otimes c)$$
and $M_x>M_y$ leads to
$$E_{l+1}(x\otimes c)=s_x-M_x<s_y-M_y=PE_{l+1}(y\otimes c),$$
which contradict the well-known fact that $P(x\otimes c\rightarrow
y\otimes c)\geq P(x\rightarrow y)$.

Now, we show that a contradiction will arise by considering the
following two cases.

Case 1: $r_n>0$. Since $1< l< nk$, we have $r_n<k$.  Then
$$M_y\geq M_x\geq x_n\alpha^{r_n}>Py_n\alpha^{r_n-1}\geq m_y$$
since $Py_n/x_n<\alpha$. Thus $M_y>m_y$, which is a contradiction.

Case 2: $r_n=0$. In this case, $x_n\leq M_x$ since $x_n$ is
included in $s_x$. By definition of $\alpha$ we have
$x_h\alpha^{k-1}<M_x$ and so $0\leq r_h\leq k-1$. Furthermore, we
can prove that $r_h>0$ since otherwise $x_h$ is in the summands of
$s_x$ and any terms not included are of the form $x_i\alpha^j$
where $1\leq i<h$. By the definition of $h$, we have
$x_i\alpha^j=Py_i\alpha^j$ for $1\leq i<h$. So the sum of the
components of $x\otimes c$ not included in $s_x$ is equal to the
sum of the components of $Py\otimes c$ not included in $s_y$. This
fact, together with the assumption that $s_x=x_y$ will lead to a
contradiction that $$1=E_1(x\otimes c)=PE_1(y\otimes c)=P.$$

Now, if $P>x_h/y_h$, then
$$m_y\leq m_x\leq x_h\alpha^{r_h-1}<Py_h\alpha^{r_h}\leq M_y$$
from our assumption that $\alpha
>x_h/(Py_h)$, and if $P<x_h/y_h$, then
$$M_y\geq M_x\geq x_h\alpha^{r_h}>Py_h\alpha^{r_h-1}\geq m_y$$
from our assumption that $\alpha >Py_h/x_h$. Again, $m_y\geq M_y$
is contradicted. That completes our proof. \hfill$\square$

\vspace{1em}

Recall that in Example 1, when $x=(0.6,0.2,0.2)$ and
$y=(0.5,0.3,0.2)$, there exists no two-dimensional useful catalyst
for the probabilistic transformation from $x$ to $y$. We show here
how to construct a higher-dimensional one by the above theorem. It
is easy to check that $$P(x\rightarrow
y)=0.8<\frac{0.6}{0.5}=\frac{x_1}{y_1},$$ so $h=1$ and we need
only take a real $\alpha$ such that $Py_3/x_3=0.8<\alpha<1$ and
$\alpha>Py_1/x_1=2/3$, that is, $0.8<\alpha<1$. In order not to
make $k$ too large, we should take $\alpha$ as small as possible.
For example, $\alpha=0.801$. Then, from the constraint $x_3>x_1
\alpha^{k-1}$ in the theorem, we have $k\geq 6$. Thus the state
$$c=(1,\alpha,\dots,\alpha^5)$$
can increase the maximal transformation probability of $x$ into
$y$.

The above theorem gives us a sufficient and necessary condition
under which the transformation $x$ into $y$ has a catalyst which
can increase the maximal probability transformation. Furthermore,
the proof process constructs a real catalyst vector. What we
should like to point out here is, however, that the catalyst
presented in the proof is not very economical in the sense that it
is usually not the minimally dimensional one among all states
which can serve as a useful catalyst. How to find a most
economical one remains for further study.

\textbf{Acknowledgement:} This work was supported by National
Foundation of Natural Sciences of China (Grant Nos: 60273003,
60321002 and 60305005) and Key grant Project of Chinese Ministry
of Education.


\begin{thebibliography}{9999}
\bibitem{BB84} C. H. Bennett and G. Brassard, Proceedings of IEEE International
Conference on Computers, Systems, and Signal Processing,
Bangalore, India, 1984 (unpublished), pp. 175¨C179.

\bibitem{BS92} C. H. Bennett and S. J. Wiesner, Phys. Rev. Lett. 69, 2881 (1992).
\bibitem{BBC+93} C. H. Bennett, G. Brassard, C. Crepeau, R. Jozsa, A. Peres, and W.
K. Wootters, Phys. Rev. Lett. 70, 1895 (1993).

\bibitem{NI99}  M. A. Nielsen, Phys. Rev. Lett. 83, 436 (1999)
\bibitem{MO79}  A. W. Marshall and I. Olkin, Inequalities: Theory of Majorization and Its Applications (Academic Press, New York, 1979). P. M.
\bibitem{Vidal99} G. Vidal, Phys. Rev. Lett. 83, 1046 (1999).
\bibitem{JP99}  D. Jonathan and M. B. Plenio, Phys. Rev. Lett. 83, 3566 (1999)
\bibitem{DK01}  S. Daftuar and M. Klimesh, Phys. Rev. A 64, 042314 (2001)
\end{thebibliography}
\end{document}